# Gerberto e l'Astronomia

# Costantino Sigismondi*

## 1. Sommario
La figura di Silvestro II, Gerbert d'Aurillac, Papa dal 999 al 1003, viene inquadrata storicamente e se ne presentano le principali opere scientifiche e didattiche in astronomia e musica. Gerberto, che ha vissuto la "mini-rinascenza" del X secolo, introdusse l'astrolabio, l'abaco e il monocordo in Europa grazie alle sue conoscenza della scienza araba, ed era il massimo esperto di acustica delle canne d'organo del suo tempo. Grazie alla sua autorità e al suo prestigio, le scienze del quadrivio (artimetica, musica, geometria e astronomia) diventarono parte del curriculum studiorum delle scuole cattedrali, e di li' a qualche secolo delle nascenti Universtà degli studi.

## Abstract
Gerbert of Aurillac was Pope Sylvester II form 999 to 1003. His history is presented in order to understand his outstanding contribution in the establishment of quadrivium sciences (arithmetics, music, geometry and astronomy) in the *curricula studiorum* of cathedral schools and therefore of forthcoming *universitates studiorum*. Gerbert allowed the first sharing of arabic scientific scientific culture (e.g. Introducing in his didactic method the astrolabium, the abacus and the monochord) with Christian world participating in person to the mini-renaissance of the tenth century.

## 2. Introduzione

A Gerberto vengono attribuite l'invenzione dell'orologio a pendolo o quello meccanico, l'introduzione in Europa delle cifre arabe incluso lo zero, insieme con la capacità di costruire organi, astrolabi. Mancando criteri storici ed il confronto con le realtà storiche a lui contemporanee, il vero e l'inverosimile sovente convivono assieme. Gli orologi meccanici appaiono in Europa tra il XIII e il XIV secolo, e le attribuzioni a Gerberto sono leggendarie. Qui esaminiamo quanto è noto su Gerberto in relazione con l'astronomia e la musica.

## 3. Gerberto in Catalogna
Seguendo Richero di Reims[1], allievo e biografo di Gerberto, sappiamo che il conte Borrell II, venuto in pellegrinaggio presso la tomba di San Geraldo, il fondatore del monastero di Aurillac, invitò il monaco Gerberto a recarsi nel 967 con lui in

---

[1] Richero, *Historia Francorum 888-995,* edito e tradotto da R. Latouche, Voll. I, II. Paris 1930, 1937. Les Classiques de l'histoire ded France au moyen age. Il testo in Latino è presente anche nella Patrologia Latina **138**, 9-170.

Catalogna per approfondire gli studi nelle matematiche con Attone, il vescovo di Vich.

Studi sugli archivi di Vich suggeriscono che Gerberto deve essere andato altrove per attingere le informazioni sull'astronomia, poiché a Vich non c'erano molti testi. La vicina abbazia di Santa Maria di Ripoll conservava a quel tempo molti testi interessanti ed era un centro di scambio culturale tra il mondo arabo e il mondo cristiano, dove si effettuavano le prime traduzioni dei testi arabi.

Presso gli arabi esistevano le traduzioni di Tolomeo e le opere dei loro astronomi, come Al-Batenio, mentre nel mondo latino i testi in greco non erano più reperibili.

Ademaro di Chabannes attribuisce a Gerberto addirittura un viaggio a Cordova, dove egli avrebbe potuto approfondire la sua conoscenza scientifica, anche a prezzo di abiurare la propria fede. Ademaro è il principale responsabile delle leggende nere su Gerberto[2], e purtroppo ancora oggi le sue dicerie circolano apocrife dovunque. Del resto a livello di opinione pubblica, ben plagiata dai media, ancora oggi, come mille anni or sono, l'ortodossia sembra inconciliabile con la conoscenza scientifica. Il problema si ripresenta sotto mentite spoglie.

## 4. L'Astronomia in Europa al tempo di Gerberto

Non solo influenze arabe nella formazione di Gerberto, ma anche la stessa regola di San Benedetto stabilisce orari precisi per la celebrazione delle ore canoniche[3].

Infatti prima di Gerberto in Astronomia in Europa erano stati trattati problemi di Cronologia, in cui personaggi come Dionigi il piccolo (532) ed Beda il Venerabile (725) avevano lavorato al fine di stabilire la data mobile della Pasqua univocamente per tutto l'orbe cattolico. Inoltre gli orari per le preghiere monastiche venivano stabiliti dalla regola di San Benedetto, e Gerberto che era benedettino li doveva conoscere molto bene, e forse ne era responsabile nel suo monastero ad Aurillac. Esistevano vari metodi o strumenti per stabilire l'ora del giorno o della notte, ed un testo di Gregorio di Tours sulle Stelle Fisse era stato scritto per aiutare i monaci in questo compito.

Infine alcuni testi di Macrobio, Marziano Capella, Boezio erano sopravvissuti dall'età classica in molti esemplari. Con la graduale scomparsa del greco nel panorama culturale europeo scomparvero anche i testi classici come quelli di Tolomeo e Aristotele che riapparirono in Europa tradotti dall'arabo nel XII secolo.

Boezio aveva cercato di fissare in latino i testi principali delle 7 arti liberali, e tra queste opere aveva scritto anche *de Institutione Musica*, che Gerberto spiega a più riprese ai suoi allievi nelle lettere[4], e il *de Astronomia*, oggi perduto, ma che Gerberto

---
[2] Si veda ad esempio: Flavio G. Nuvolone, *Gerberto d'Aurillac-Silvestro II visto da Ademaro di Chabannes,* in F. G. Nuvolone ed. op. cit. 2001 p. 599-657. Si veda anche Huguette Taviani-Carozzi, *An Mil et Millénarisme: le Chronicon d'Adémar de Chabannes,* ibidem 779 -821.
[3] Cfr. S. Benedetto da Norcia, *Regola*, Padri Benedettini di Subiaco, Subiaco 2001, capitoli 8 e 47.
[4] H. PRATT LATTIN, *The Letters of Gerbert with his papal privileges as Sylvester II*. Translated with an Introduction by Harriet PRATT LATTIN, Columbia University Press, New York, (1961).Lettera 4 e Lettera 5 entrambe a Costantino

stesso aveva rinvenuto in una sua trasferta a Mantova, mentre era abate di Bobbio[5]. Gerberto è stato giudicato anche il massimo bibliofilo del medioevo[6].

## 5. Gerberto astronomo

La lettera di Gerberto a Lupitus (Lopez) di Barcellona, datata 984[7], ci mostra lo scolastico di Reims desideroso di seguire gli sviluppi in Catalogna delle conoscenze dell'astronomia araba. "Bubnov scoprì nella Bibliothèque Nationale de Paris il frammento di un trattato sull'astrronomia tradoto ed adattato da fonti arabe: Duhem[8] ritiene che l'autore di un trattato pervenuto integro fino a noi, il *Liber de astrolabio*, si sia servito appunto di tale frammento per comporre la sua opera, e formula l'ipotesi che possa trattarsi d'un frammento del trattato richiesto da Gerberto a Lopez. L'attribuzione del *Liber de astrolabio* a Gerberto non è incontestabile, ma Duhem la considera quasi certa"[9]. Passata l'epoca in cui si smontavano sistematicamente i grandi personaggi storici, oggi gli studiosi hanno accettato che Gerberto sia stato l'autore di questo trattato[10].

Gerberto scrisse anche un altro trattato, *Sul calcolo con l'abaco*, derivato anch'esso da fonti arabe. Gerberto riscosse la devota ammirazione dei suoi contemporanei grazie agli strumenti astronomici che aveva acquistato in Spagna ed alla sua conoscenza della matematica e della scienza. Secondo Duhem, servendosi di fonti arabe per il suo libro sull'astrolabio egli creò una nuova moda. Durante l'undicesimo secolo apparvero parecchi trattati latini sugli strumenti astronomici, che seguivano fedelmente modelli arabi. Un imitatore di Gerberto che merita di essere ricordato, secondo Duhem, fu Hermann di Reichenau (1013-54), che scrisse sull'astrolabio e sull'abaco. Bisogna rammentare inoltre, di passaggio, la trattazione sull'astrolabio di Ugo di San Vittore (m. 1141), che ne discute gli usi nei primi capitoli della parte pratica della sua *Pratica geometriae*[11]. Gerberto fu famoso anche per la quantità di manoscritti che si procurò per la sua biblioteca: è stato detto di lui che fu il maggior collezionista di libri del Medioevo. La sua raccolta comprendeva una copia della traduzione dell'*Introduzione all'Aritmetica* di Nicomaco eseguita da Boezio, e la cosiddetta "*Geometria di Boezio*", una compilazione del secolo XI che rispecchia solo vagamente l'opera autentica andata perduta. Il libro sulla Geometria scritto da Gerberto dimostra una notevole familiarità con la traduzione di Boezio

---

di Fleury.
[5] Lettera 15 all'Arcivescovo Adalberone di Reims.
[6] Vedasi M. Oldoni, *Silvestro II,* Enciclopedia dei Papi, Treccani vol. II 2000, e W. H. Stahl op. cit. p. 316.
[7] Lettera 32, numerazioni dell'edizione in inglese dalla Harriet PRATT LATTIN nel 1961 (*The Letters of Gerbert with his papal privileges as Sylvester II*. Translated with an Introduction by Harriet PRATT LATTIN, Columbia University Press, New York, 1961).
[8] P. Duhem, *Le Système du Monde*, Paris 1913-1959, vol. III pp. 164-5.
[9] W. H. Stahl, *La Scienza dei Romani,* Laterza, Bari 1974 pp. 315-6.
[10] Uta Lindgren, "Représentatn de l'age obscur ou à l'aube d'un essor? Gerbert et les Arts Liberaux" in "Gerberto d'Aurillac da Abate di Bobbio a Papa dell'Anno 1000" Atti del Congresso Internazionale Bobbio, 28-30 Settembre 2000, editi da F. G. Nuvolone.
[11] Nel capitolo 42 ipotizza la possibilità di determinare la distanza del Sole per mezzo di osservazioni effettuate da due punti notevolmente distanti: si tratta di un procedimento greco-arabo. Ma poi Ugo di San Vittore include la puerile discussione sulle dimensioni dell'orbita solare e sulle grandezze relative della Terra e del Sole.

dell'*Introduzione* nicomachea, ma è così diversa dagli *Elementi* di Euclide che l'autore non può aver conosciuto quest'opera nella forma boeziana."[12]

La "mini-rinascenza" della fine del secolo X[13] ha il suo apice negli anni successivi alla permanenza di Gerberto in quelle terre. Tuttavia Gerberto ne era al corrente anche da Reims, ciò che dimostra come la rete di comunicazione scientifica fosse efficiente anche nei presunti secoli bui.

Un altro pregiudizio da evitare nell'accostarci a figure così lontane nel tempo è quello di ritenere le nostre argomentazioni come gli unici sprazzi di luce in una realtà che incombesse imponderabile su inermi protagonisti. Mai come in Gerberto vediamo un uomo attivo a livello europeo sul piano ecclesiale, diplomatico e scientifico, e quando questi fu eletto papa, la sua fama era già universale, tanto che nei documenti ufficiali si firmava *Silvestro, Gerberto, Vescovo Romano*[14], oppure Silvestro, *che è anche Papa Gerberto*[15].

Anche da papa Gerberto si rivolse ancora al mondo catalano[16].

Per l'astronomia abbiamo una lettera[17] inviata a Costantino di Fleury dove descrive la costruzione della sfera ruotante, per uso didattico, che replica il comportamento della sfera delle stelle fisse.

In un'altra Gerberto lettera presenta le zone climatiche della terra[18]: un argomento di geografia diremmo noi oggi, ma allora questo tipo di geografia era astronomica. Ad ogni zona climatica corrisponde un valore della lunghezza del giorno più lungo dell'anno.

## 6. Gerberto matematico

Gerberto scrisse anche un altro trattato, *Sul calcolo con l'abaco*, derivato anch'esso da fonti arabe. Gerberto fu famoso anche per la quantità di manoscritti che si procurò per la sua biblioteca: è stato detto di lui che fu il maggior collezionista di libri del Medioevo. La sua raccolta comprendeva una copia della traduzione dell'*Introduzione all'Aritmetica* di Nicomaco eseguita da Boezio, e la cosiddetta "*Geometria di Boezio*", una compilazione del secolo XI che rispecchia solo vagamente l'opera autentica andata perduta. Il libro sulla Geometria scritto da Gerberto dimostra una notevole familiarità con la traduzione di Boezio dell'*Introduzione* nicomachea, ma è così diversa dagli *Elementi* di Euclide che l'autore non può aver conosciuto quest'opera nella forma boeziana."[19]

---

[12] W. H. Stahl, *La Scienza dei Romani,* Laterza, Bari 1974 pp. 315-6.
[13] Cfr. Michel Zimmermann, *La Catalogne de Gerbert*, in Gerbert *l'Européen,* Charbonnel-Iung éd. *Gerbert l'Européen,* Actes du colloque d'Aurillac, mémoire de la société –La haute Auvergne 3, Aurillac 1997, p. 86.
[14] Lettera 251,
[15] Lettera 264.
[16] Lettera 255.
[17] Lettera 2.
[18] Lettera 161, al Frate Adamo.
[19] W. H. Stahl, *La Scienza dei Romani,* Laterza, Bari 1974 pp. 315-6.

## 7. Gerberto musico

Gerberto era anche il massimo esperto di organi e di organaria del secolo X, l'attribuzione del trattato de Mensura fistularum è stata provata solo nel 1970, ma ancor oggi questo concetto non è ancora entrato nei manuali di storia della musica. Diversi organi erano attribuiti a lui.
Klaus Jurgen Sachs[20] ha trovato nella biblioteca di Madrid un manoscritto[21] più antico di quello della biblioteca Vaticana, dove invece l'opera era attribuita a Bernellino, e cosi' pubblicata da Gerbert Martin nel 1784 nel suo Scriptores Ecclesiastici de Musica Sacra. Nel manoscritto di Madrid l'attribuzione è a Gerberto, che era stato riconosciuto come esperto senza eguali in musica e organo anche da Papa Giovanni XIII oltre che dai suoi allievi che lo continuano ad interpellare in questioni di musica teoriche e pratiche. Alla corte di Ottone III, nella cappella imperiale, Gerberto si occupò di musica[22].

## 8. Gerberto docente

L'esistenza di un trattato sulla Musica, di quello *de Utilitatibus Astrolabii*, di quello sulla Geometria ci testimoniano un docente che, pur tra i suoi numerosi impegni diplomatici ed ecclesiastici, ha provveduto a scrivere dei testi per i suoi studenti, e non solo risposte brevi su lettera a precise domande.
Anche se Geberto è rimasto celebre per la sua brevità, e il carattere diretto delle sue lettere, nonostante fosse il più famoso retore del suo tempo.

Nel corso di questo intervento vi mostro in stile gerbertiano l'effetto del moto dei cieli mediante la montatura equatoriale del telescopio ed un fascetto laser.
La grande novità introdotta dal docente Gerberto d'Aurillac fu quella di mettere in mano agli studenti degli strumenti così che essi potessero toccare con mano la nozione che stavano studiando e lo stesso Richero di Reims, allievo di Gerberto, dichiarò che in questo modo il funzionamento della sfera celeste restò loro impresso in modo indelebile[23].

## Conclusioni

---

[20] *"Mensura Fistularum", Die Mensurierung der Orgelpfeifen im Mittelalter*, Stuttgart 1970
[21] Biblioteca Nacional, Ms 9088, f. 125-128
[22] Pratt Lattin op. cit., p. 16.
[23] "Il calcolo di questo strumento era così accurato che con il suo diametro [asse] puntato al polo e il suo semicerchio [equatore] rivolto verso il cielo, i cerchi [celesti] sconosciuti venivano portati alla luce e stampati profondamente nella memoria." Richero, *Historia Francorum*, **III**, 51.

Con l'incontro di oggi ci siamo proposti di ripresentare la figura di Silvestro II, Gerbert d'Aurillac, con l'occasione di questo giorno millenario, curando anche la diffusione a larga scala degli studi gerbertiani, avendo aperto un sito internet universitario per lo scopo.

Queste conferenze verranno edite in un CD-ROM in cui sara' presente anche il video della dimostrazione condotta in aula.

Gli studi medievali hanno giovato grandemente della figura di Gerberto per ricostruire il periodo del X secolo, così povero di fonti. La storia della scienza e della tecnica vede sempre più in Gerberto una figura chiave per descrivere il primo incontro tra il mondo arabo e quello cristiano sul piano culturale, così come il massimo sviluppo della scienza latina medievale nel campo della musica. Per queste ragioni a questa giornata gerbertiana seguirà l'anno venturo un convegno di più ampio respiro proprio nel giorno anniversario del 12 maggio, che ci auguriamo possa svolgersi poi a cadenze regolari.

Mi auguro che a voi questo Costantino, allievo di Gerberto di 1000 anni dopo, possa aver chiarito qualcosa in più del moto dei cieli e dell'armonia degli organi.

**Bibliografia**